\pdfoutput=1

\documentclass[11pt]{article}

\usepackage{acl}
\usepackage{geometry}
\usepackage{times}
\usepackage{latexsym}
\usepackage{amsmath}
\usepackage{bm}

\usepackage{booktabs}

\usepackage[T1]{fontenc}

\usepackage[utf8]{inputenc}

\usepackage{microtype}

\usepackage{inconsolata}

\usepackage{graphicx}
\usepackage{xspace}

\newcommand{\method}{\textbf{{TheoremLlama}}\xspace}

\usepackage{listings}
\usepackage{amssymb}
\usepackage{array} 

\usepackage{color}
\definecolor{keywordcolor}{rgb}{0.7, 0.1, 0.1}   
\definecolor{tacticcolor}{rgb}{0.0, 0.1, 0.6}    
\definecolor{commentcolor}{rgb}{0.4, 0.4, 0.4}   
\definecolor{symbolcolor}{rgb}{0.0, 0.1, 0.6}    
\definecolor{sortcolor}{rgb}{0.1, 0.5, 0.1}      
\definecolor{attributecolor}{rgb}{0.7, 0.1, 0.1} 

\title{TheoremLlama: Transforming General-Purpose LLMs into Lean4 Experts}

\author{
 \textbf{Ruida Wang\textsuperscript{1}}\thanks{First authors},
 \textbf{Jipeng Zhang\textsuperscript{1}}\footnotemark[1],
 \textbf{Yizhen Jia\textsuperscript{1}}\footnotemark[1],
 \textbf{Rui Pan\textsuperscript{2}},
\\
 \textbf{Shizhe Diao\textsuperscript{3}},
 \textbf{Renjie Pi\textsuperscript{1}},
 \textbf{Tong Zhang\textsuperscript{2}},
\\
 \textsuperscript{1}Hong Kong University of Science and Technology, \\
 \textsuperscript{2}University of Illinois Urbana-Champaign, 
 \textsuperscript{3}NVIDIA
\\
 \small{
   \textbf{Correspondence:} 
    \href{rwangbr@connect.ust.hk}{rwangbr@connect.ust.hk}, 
    \href{jzhanggr@ust.hk}{jzhanggr@ust.hk}, 
    \href{yizhen.jia96@gmail.com}{yizhen.jia96@gmail.com},  
 } \\
 \small{
    \href{ruip4@illinois.edu}{ruip4@illinois.edu}, 
    \href{sdiao@nvidia.com}{sdiao@nvidia.com}, 
    \href{rpi@ust.hk}{rpi@ust.hk}, 
    \href{tongzhang@tongzhang-ml.org}{tongzhang@tongzhang-ml.org}
 }
}

\begin{document}
\maketitle
\begin{abstract}
Proving mathematical theorems using computer-verifiable Formal Languages (FL) like Lean significantly impacts mathematical reasoning. One approach to formal theorem proving involves generating complete proofs using Large Language Models (LLMs) based on Natural Language (NL) proofs. 
However, due to the scarcity of aligned NL and FL theorem-proving data, most modern LLMs exhibit suboptimal performance.
This scarcity results in a paucity of methodologies for training LLMs and techniques to fully utilize their capabilities in composing formal proofs. 
To address these challenges, this paper proposes \method, an end-to-end framework that trains a general-purpose LLM to be a Lean4 expert. \method includes NL-FL dataset generation and bootstrapping method to obtain aligned dataset, curriculum learning and block training techniques to train the model, and iterative proof writing method to write Lean4 proofs that work together synergistically.
Using the dataset generation method in \method, we provide \textit{Open Bootstrapped Theorems} (OBT), an NL-FL aligned and bootstrapped dataset. Our novel NL-FL bootstrapping method, where NL proofs are integrated into Lean4 code for training datasets, leverages the NL reasoning ability of LLMs for formal reasoning. 
The \method framework achieves cumulative accuracies of 36.48\% and 33.61\% on MiniF2F-Valid and Test datasets respectively, surpassing the GPT-4 baseline of 22.95\% and 25.41\%. Our code, model checkpoints, and the generated dataset is published in \href{https://github.com/RickySkywalker/TheoremLlama}{GitHub}

\end{abstract}
\section{Introduction}\label{sec:intro}

The ability to perform logical reasoning has always been regarded as a cornerstone of human intelligence and a fundamental goal of machine learning systems~\cite{newell1956logic}. 
Among these tasks, mathematical reasoning is considered crucial for evaluating the capabilities of Large Language Models (LLMs).
However, in modern mathematics, verifying the correctness of theorem proofs written in natural language is challenging, complicating the assessment of LLMs' mathematical reasoning in advanced topics. 
Additionally, the rapid development of modern mathematics and the increasing complexity of proofs pose significant barriers to reviewing their correctness.
This has led to erroneous proofs that require considerable effort to be identified by the mathematical community, as exemplified by the process of proving Fermat's Last Theorem \cite{taylor1995ring}.
To address these issues, formal mathematical languages such as Lean \cite{de2015lean, moura2021lean}, Isabelle \cite{paulson1994isabelle}, and HOL Light \cite{harrison2009hol} have been developed.
These languages allow computers to automatically verify proofs, providing a clear standard for evaluating mathematical theorem proofs and significantly impacting both the mathematical and computer science communities.

\begin{figure*}[t]
    \vspace{-0.25in}
    \centering
    \includegraphics[width=0.99\linewidth]{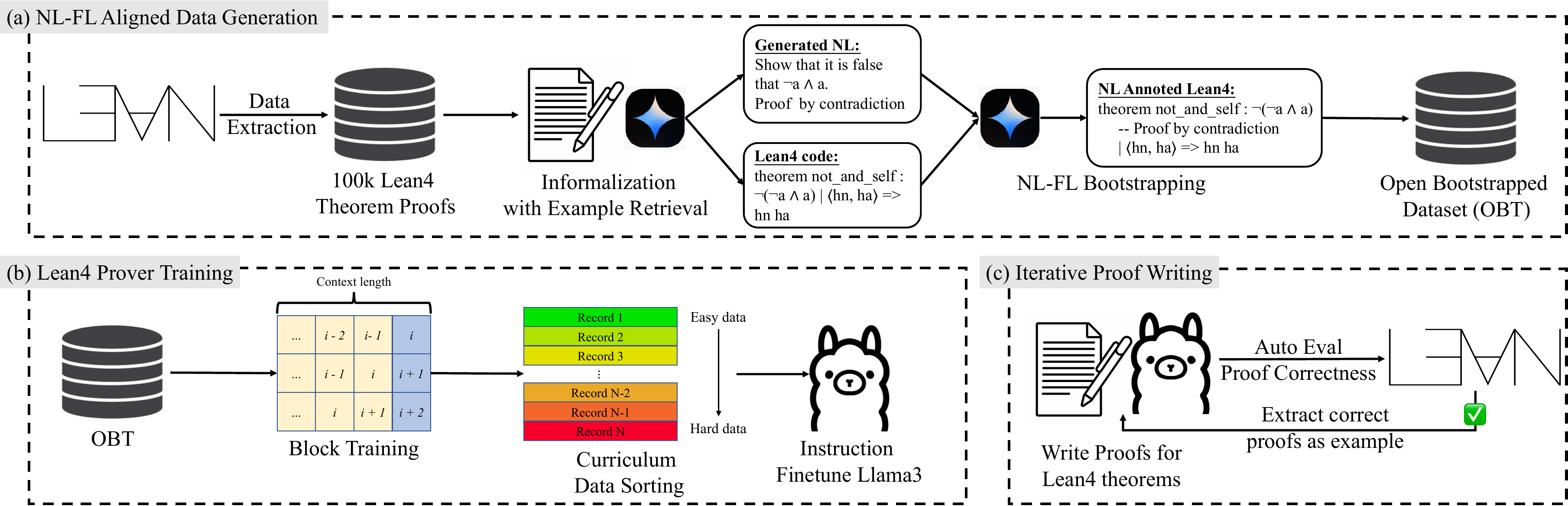}
    \caption{
    \method Framework: (a) NL-FL Aligned Data Generation: We first extract Lean4 data from Mathlib4. Subsequently, we fine-tune a T5 encoder to search for the best examples to guide the informalization of the extracted data. Then, we apply Gemini-1.5 to informalize extracted theorems with retrieved examples. Finally, we perform NL-FL Bootstraping to integrate natural language reasoning into Lean4 codes. Using this generation method, we have the OBT dataset. (b) Lean4 Prover Training: We use block training to enhance the in-context ability and the curriculum data sorting to let LLM learn from easy to hard data. These techniques can make LLM better learn unfamiliar Lean4 theorem proving tasks. (c) Iterative Proof Writing: We iteratively use the correct proofs from the same dataset of the previous iterations as in-context examples to enhance the proof-writing ability of the LLM.
    }
    \label{fig:main}
    \vspace{-0.2in}
\end{figure*}

However, writing mathematical proofs in Formal Language (FL) requires significant expertise and effort. 
Additionally, formal proofs involve much repetitive and tedious work~\cite{jiang2022draft}, which is not customary for mathematicians who are more familiar with high-level proofs.
Consequently, there has been significant demand for automated theorem-proving using FL, leading to a considerable number of works on this task~\cite{polu2020generative, polu2022formal, jiang2021lisa, jiang2022thor, jiang2022draft, yang2024leandojo}. 
However, most of these works rely on searching methods in an infinite space of possible tactics to complete the proof, resulting in unaffordable computational costs (e.g., \citet{polu2022formal} uses 2,000 A100 GPU hours for training) in finishing complex proofs and not fully leveraging NL reasoning ability of LLMs.
Recent advancements in LLMs, especially in reasoning~\cite{wei2022chain} and coding~\cite{roziere2023code}, have prompted researchers to explore using them to write formal proofs guided by natural language~\cite{wu2022autoformalization, jiang2022draft}.

In this paper, we focus on enabling general-purpose LLMs to write formal Lean4 proofs guided by natural language (NL) proofs. We have chosen Lean4 because it has recently garnered considerable attention from the mathematical community~\cite{Tao2023, Tao2023b, avigad2020mathematics}, whereas Lean3 and Isabelle are older formal languages. Despite the potential demonstrated by previous works~\cite{wu2022autoformalization, jiang2022draft} in similar tasks using Isabelle, the few-shot performance of LLMs in Lean4 remains relatively unsatisfactory in Lean 4 (details in Appendix~\ref{appendix:pretrain}). This is because Lean4 is a more concise FL that differs significantly from NL, making the direct transfer of reasoning abilities from NL to Lean4 infeasible. The situation is exacerbated by the inclusion of confusing Lean3 code in the LLMs' pre-training data (details in Appendix~\ref{appendix:pretrain}). More importantly, there is a significant lack of NL-FL aligned data, making the training of LLMs to write Lean4 proofs an overlooked and challenging task. Additionally, \citet{jiang2022draft} indicates that there remains a large potential for researchers to fully utilize LLMs in writing formal proofs.

To address these challenges, we propose \method, an end-to-end framework that transforms a general-purpose LLM into a Lean4 expert. The framework overview is presented in Fig.~\ref{fig:main}. Our framework comprises three major components that work synergistically:

(a) NL-FL Aligned Data Generation: This component tackles the data scarcity problem. During generation, we identified Mathlib4, a pure Lean4 repository containing 100k proofs of important mathematical theorems. We informalize Mathlib4 (i.e., write natural language theorem statements and proofs based on Lean4 code) using a Gemini-1.5 with retrieved examples from a fine-tuned T5 encoder. Subsequently, we bootstrap the NL-FL aligned data by integrating the natural language proofs into Lean4 code via comments. This process of embedding natural language reasoning within the formal language code helps the LLM better understand the theorems and leverages its natural language reasoning ability to perform formal reasoning. Following this generation and bootstrapping method, we create the \textit{Open Bootstrapped Theorems} (OBT) dataset.

(b) Lean4 Prover Training: This component introduces training methods that are currently understudied in the field. It includes block training techniques to improve the LLM's in-context learning ability and curriculum data sorting tactics to ensure a smoother training process. Using this method, we fine-tune Llama3-8B-Instruct to be a Lean4 expert with the OBT dataset.

(c) Iterative Proof Writing: This component enhances the LLM's ability to write formal proofs by using previously generated correct proofs as in-context examples to further improve its formal reasoning capabilities.

We summarize our contributions in this paper as follows: (1) We propose \method, an end-to-end framework that transforms a general-purpose LLM into a formal proving expert. 
\method spans from NL-FL aligned dataset generation to Lean4 prover training techniques and iterative proof writing for Lean4 prover. It amends the significant data scarcity problem by contributing to the OBT dataset. Additionally, it contains LLM training and proof writing methods that have largely been overlooked in Lean4 theorem proving. 
(2) Our major innovative point is the NL-FL bootstrapping method, which integrates informal proofs into Lean4 code. this method enhances the LLMs' abilities by using training data to transfer their informal reasoning capabilities to Lean4 proof writing. 
(3) We conduct extensive experiments using \method, which achieves 36.48\% and 33.61\% accuracy rate on MiniF2F-Valid and Test, largely suppressing GPT-4 baseline (25.41\% and 22.95\% separately). Additionally, we perform a thorough ablation study to prove the effectiveness of major components in dataset generation and training.

Furthermore, we open-source the OBT dataset, model checkpoints, and codes to support further research in the community. Under a reasonable GPU footprint for \method (the fine-tuning only takes about 32 hours for an 8 GPU A6000 machine) our work will significantly lower barriers to academic researchers in corresponding fields of obtaining considerably well-behaved Lean4 prover.

\section{Methodology}\label{sec:meth}

In this section, we present the details of \method, including generation methods for the OBT dataset. The key idea for our framework is to enable LLMs to perform well in the unfamiliar Lean4 theorem proving task under the circumstances of limited or even confusing data during its pre-training. We introduce the Dataset Generation method in Section \ref{meth:dataGen}, illustrate the training techniques for Lean4 prover training using the OBT dataset in Section~\ref{meth:train}, and propose an Iterative Proof Writing method LLM prover in Section~\ref{meth:iter}. 
The task that this methodology works on can be defined as: "Training the general purpose LLM to be an expert in Lean4 whole-proof generation under the guidance of Natural Language Proofs." 

\subsection{NL-FL Aligned Data Generation}\label{meth:dataGen}
This section describes the Natural Language (NL) - Formal Language (FL) Aligned Dataset Generation method. As previously discussed, we chose Lean4 as the formal language for our study. 
The dataset generation aims to enhance the LLM's ability in Theorem proving from the dataset point-of-view. To the best of our knowledge, no open-source Lean4 NL-FL aligned dataset exceeds 1k records, our dataset generation provides \textit{Open Bootstrapped Theorems} (OBT) dataset containing 106,852 NL-FL aligned and bootstrapped theorems.

\subsubsection{Lean4 Proof Extration}\label{dataGen:extract}

Although there is no NL-FL aligned dataset, for Lean4, there is Mathlib4, a repository containing 100k high-quality, human-crafted proofs. It is a general repository that contains the most important definitions and theorems from basic mathematics, including logic, set theory, number theory, and algebra; to advanced topics like topology, differential geometry, and real/complex analysis. Mathlib4 offers the LLM a high-quality, and comprehensive foundation for dataset generation tasks. Previous works have used such a dataset for tree-search prover training \cite{yang2024leandojo, polu2022formal}. We directly extract Mathlib4's theorems from the LeanDojo~\cite{yang2024leandojo} repository for further dataset generation.

\subsubsection{Informalization with Example Retrival}\label{dataGen:deform}

To the best of our knowledge, the potential for using Mathlib4 as a dataset for training LLMs to generate formal proofs based on natural language guidance is an understudied field. This is due to Mathlib4 does not contain corresponding natural language statements for most of the theorems. However, with the development of modern LLMs, we propose a way to generate the NL-FL aligned dataset for Mathlib4. This method, which writes informal proofs from formal proofs, is called formalization

Due to the mix of Lean4 and Lean3 data on the internet, LLMs pre-trained on web-scale data only have limited ability to recognize Lean4 proofs and may be interfered by perplexing Lean3 data (Appendix~\ref{appendix:pretrain}.)
Therefore, it is important to have high-quality in-context examples for informalization. Inspired by the idea of contrastive loss~\cite{izacard2021unsupervised}. We develop the example retrieval method to extract such high-quality examples. The first step for our example retrieval is using the Natural Language annotated MiniF2F dataset \cite{jiang2022draft} to fine-tune the ByT5-Tacgen model provided by \citet{yang2024leandojo}, which trained on pure Lean4 data, has a relative good understanding of both NL and FL data.

Specifically, we fine-tune the ByT5 encoder to align the cosine similarity of the theorem statement of natural language and Lean4 code. The sentence-level encoding is obtained by mean pooling of every token's encoding. To prevent the model from producing trivial results, we add in-batch negative sampling in the loss function. Thus, the loss for fine-tuning ByT5 is:
\[
\begin{aligned}
\vspace{-0.1in}
    \mathcal{L} =& 1 - \cos (\bm{x}_{NL}, \bm{x}_{FL}) + \\
                 & \frac{1}{2} (\cos (\bm{x}_{NL}^{(-)}, \bm{x}_{FL}) + \cos (\bm{x}_{NL}, \bm{x}_{FL}^{(-)}))
\end{aligned}
\vspace{-0.1in}
\]
where $\bm{x}_{NL/FL}$ represents sentence encoding; $\bm{x}^{(-)}$ means not aligned NL/FL statement in the same batch as the negative sample.

Subsequently, we use this encoder to encode the Lean4 theorem statement in Mathlib4 and the informal theorem statement in a tiny NL-FL aligned dataset (less than 100 theorem proofs, provide by \citet{yang2024leandojo}). 
We select a few examples with the highest similarity and use these as in-context examples to query Gemini-1.5 \cite{reid2024gemini} to obtain informalized theorem statements and proofs for the theorems in Mathlib4.

After informalization, we conduct a primary-level data quality check. \citet{wu2022autoformalization} found that most of the informalization made by LLMs generally makes sense so our quality check majorly focuses on removing abnormal behavior of LLMs, including repeated generation, overlength generation, and other erroneous data. We iteratively query the Gemnin-1.5 with failed examples, and ultimately, we obtain an NL-FL aligned dataset consisting of 106,852 theorems, a much larger dataset than any currently open-sourced NL-FL aligned dataset for Lean4.

\subsubsection{NL-FL Bootstrapping}\label{dataGen:bootstrap}

We find that due to the significant differences between performing natural language reasoning and Lean4 theorem proving, externally NL-guided training data is not sufficient to enable LLMs to develop strong Lean4 theorem-proving abilities.
It is common for the LLMs to lose track of the proof and repeatedly generate the final Lean4 tactic. 
Inspired by findings in LLM coder \cite{song2024code}, where NL comments of code task description can largely improve the performance of LLM coders. We propose the novel NL-FL Bootstrapping. This is a simple but effective method to enhance the LLMs' Lean4 proof writing ability by integrating natural language reasoning into Lean4 proofs in Mathlib4. 

We achieve such an integration by providing Gemini with NL and FL of the theorem and asking it to document the natural language proof to the Lean4 code through comment. We ensure the correctness of the bootstrapped data by running a check algorithm that removes all comments in the generated code and makes sure it is the same as the original code.

This bootstrapping approach aims to lower the barrier between complex and unfamiliar-to-LLM Lean4 formal language reasoning and natural language reasoning.
We find that most modern LLMs possess relatively strong natural language reasoning abilities but lack familiarity with formal reasoning. 
This method helps LLMs transfer their natural language reasoning skills to Lean4 theorem proving by bootstrapping the dataset, prompting the LLM to perform both formal and informal reasoning simultaneously. LLMs trained with the bootstrapped dataset will learn to better utilize the NL steps to guide Lean4 proof writing. Following above generation and bootstrapping method, we have the \textit{Open Bootstrapped Theorems} (OBT) dataset for training LLMs.

\subsection{LLM Prover Training}\label{meth:train}

Training LLMs to generate whole proof based on natural language guidance is an under-explored field of study due to the lack of datasets. There are only a few studies that discuss the training method of the LLM for such a task. This section proposes two instruction fine-tuning techniques to train the LLMs for formal reasoning tasks, namely Block Training and Curriculum Data Sorting. 

\subsubsection{Block Training}\label{train:block}

The Block Training method aims to incorporate the in-context learning ability during training. For standard instruction fine-tuning in formal theorem proving, we use natural language as the input and the corresponding Lean4 with bootstrapped NL reasoning as the target output to fine-tune the LLM. In the Block Training, we view the tokenized training dataset as a ring of text. We take full advantage of the context length of LLM by filling it with examples of previous records. Formally, the original training data for $i$-th record is:
\vspace{-0.1in}
\[\{ \text{"Instruction": } NL_i, \text{"Target": } FL_i \}
\vspace{-0.1in}\]
where $NL_i$ is the natural language of $i$-th record and $FL_i$ is its corresponding bootstrapped Lean4 code. After Block Training, the $i$-th data record is:
\vspace{-0.1in}
\[\begin{aligned}
    \{ & \text{"Instruction": } ``NL_{i - k}, FL_{i - k}; \cdots FL_{i - 1}; NL_i'', \\
       & \text{"Target": } ``FL_i''\}
\end{aligned}
\vspace{-0.1in}\]

where $k$ is the number of examples that just fill the context length.

Using the block training method, we enhance the LLM's in-context learning ability for Lean4, providing a better understanding of examples when writing proofs.

\subsubsection{Curriculum Data Sorting}\label{train:learn}

Because modern LLMs have limited exposure to writing Lean4 proofs with NL guidance during pre-training, they are unfamiliar with this task. This issue is evident as LLMs with a significant difference in parameters show only slight performance differences in these tasks (details in Section~\ref{exp:results}). Inspired by previous work in Curriculum Learning~\cite{polu2020generative, soviany2022curriculum}, we propose a training data sorting technique named Curriculum Data Sorting to enable LLMs to learn this unfamiliar task from easy to difficult.

Specifically, we reorganize the generated training dataset by difficulty level. We measure the difficulty of a Lean4 proof by the steps it takes to solve all goals and sort the training data records with easier data at the beginning and harder data at the end. 
This sorting method allows the LLM to first learn to solve trivial and easy problems before tackling complex proofs. It largely stabilizes the loss curve of training and improves the performance of the LLM prover.

\subsubsection{Instruction Fine-tuning}\label{train:finetune}

Using Blocked Training and Curriculum Data Sorting on the OBT dataset, we perform the Instruction Fine-tuning on Llama3-8B-Instruct using SFT trainer in an autoregressive manner. The given instruction is a natural language statement and proof of a theorem, along with a Lean4 theorem statement, and use examples in the dataset filling context windows. The target output is the Lean4 proof bootstrapped with natural language explanations. This process trains the LLM to leverage its natural language reasoning ability to write Lean4 proofs.

\subsection{Iterative Proof Writing}\label{meth:iter}

\citet{jiang2022draft} have demonstrated that the potential of LLMs in formal reasoning is largely undervalued. Typically, LLMs possess relevant knowledge but lack appropriate methods to extract this knowledge in Lean4 form. To further harness the LLM's ability to prove Lean4 theorems, inspired by \citet{wang2023let}, we propose the Iterative Proof Writing strategy. This method involves initially having the prover finish as many theorems in a dataset as possible. Then, use the Lean-verified correct proofs written by the current prover as additional examples for proof writing in the next iteration. The iteration stops when the maximum number of steps is reached or no additional theorems can be proved with the given examples.  
This step is effective because there are potential distribution shifts between the generated and the real-world natural language statement and proof, using examples from the same dataset, such differences can be largely mitigated.

\section{Experiments}\label{sec:exp}

\begin{table*}
    \centering
    \vspace{-0.25in}
    \small
    \begin{tabular}{ccccc}
        \toprule
        \textbf{Method}                 & \textbf{Model size} & \textbf{MiniF2F-Valid} & \textbf{MiniF2F-Test}  & \textbf{Average} \\
        \midrule
        \textit{Tree-search Methods} \\
        \textbf{Expert Iteration}       & 774M                 & 28.5\%                & 25.9\%             & 27.2\% \\
        \textbf{ReProver}               & 229M                 &  -                    & 25.00\%            & - \\
        \midrule
        \textit{Unfinetuned LLM} \\
        \textbf{GPT-4-Turbo}            & > 1T                 & 25.41\%               & 22.95\%            & 24.18\% \\
        \textbf{Gemini-1.5-pro}         & -                    & 29.92\%               & 27.87\%            & 28.90\% \\
        \textbf{Llama3-Instruct}        & 8B                   & 25.41\%               & 20.08\%            & 22.75\% \\
        \midrule
        \textit{Math Expert LLM} \\
        \textbf{Llemma}                 & 31B                  & 21.03\%               & 22.13\%            & 21.58\% \\      
        \textbf{DeepSeek-Math}          & 7B                   & 25.80\%               & 24.60\%            & 25.20\%\\
        \midrule
        \textbf{\method}  & 8B                   & \textbf{36.48\%}      & \textbf{33.61\%}   & \textbf{35.04\%} \\
        \bottomrule
    \end{tabular}
    \caption{Main experimental results. Each LLMs result takes 128 rounds of generation, \method are cumulative results for multiple iterations of proofs}
    \label{tab:main}
    \vspace{-0.2in}
\end{table*}

We conduct extensive experiments on the MiniF2F-Lean4 dataset \cite{zheng2021minif2f} to test the effectiveness of \method framework on formal reasoning with NL guidance. We also conduct ablation studies (Section~\ref{exp:ablation}) and case studies (Section~\ref{exp:case}) to further validate the \method.

\subsection{Experiment Setup}\label{exp:steup}

\subsubsection{Dataset and Task}\label{setup:dataset}
In this work, we evaluate the \method Lean4 formal reasoning ability on MiniF2F-Test and Validation dataset \cite{zheng2021minif2f} and NL theorem statement and proofs provided by \citet{jiang2022draft}. We contribute the Lean4 version of MiniF2F-Test based on~\cite{yang2024leandojo}. MiniF2F is a standard testing dataset for evaluating the performance of formal provers. Both the test and validation datasets contain Lean4 statements of 244 problems. The range of problems varies from high-school competition questions to undergraduate-level theorem proofs. Specifically, MiniF2F comprises a total of 488 problems from three sources: (1) 260 problems sampled from the MATH dataset~\cite{hendrycks2021measuring}; (2) 160 problems from high-school mathematical competitions (including AMC, AIME, and IMO); (3) 68 manually crafted problems at the same difficulty level as (2). 
We are unable to use the Mathlib dataset for comparison because some baselines does not open-source their train-test split of such dataset.

Our task is to query LLM to generate the complete Lean4 proofs for the mathematical problems in MiniF2F based on their Lean4 statement and NL statement and proof together using no more than 16 in-context examples. All the imports are manually set to lighten the workload of LLM.

\subsubsection{Baseline}\label{setup:baseline}

Due to the lack of previous studies that use LLMs to generate complete proofs for Lean4; and most of the existing works working on Reinforcement Learning (RL) or searching methods, there are no universally approved baselines for comparison.
Many existing works are focusing on Isabelle~\cite{jiang2022draft, jiang2022thor, wu2022autoformalization}, a language that is largely different from Lean4, making direct comparison infeasible~\cite{yang2024leandojo}. 
Many Lean-based methods concentrate on online iteration with Lean \cite{lample2022hypertree, polu2022formal}.

Therefore, our baseline selection focuses on tree-based methods without RL and few-shot LLM proof writing. 
The baselines we use include: 
(1) Expert Iteration~\cite{polu2022formal}: A tree search method based on GPT-f~\cite{polu2020generative} that applies expert iteration to enhance the performance\footnote{Since full training of such methods uses closed source model and full training of such models takes more than 2,000 A100 GPU hours, for a fair comparison, we use $\theta_1$ result as baseline}; 
(2) ReProver~\cite{yang2024leandojo}: The Lean4 tree-search baseline that builds on ByT5 to search for tactics based on current formal state and goal. 
(3) Few-shot LLMs: This baseline focuses on directly querying LLMs to get the full proof of a formal theorem in a few-shot manner. In particular, we choose GPT-4-Turbo~\cite{achiam2023gpt}\footnote{From \cite{Bambhaniya2024}, we infer the parameter size of GPT-4 Trubo to be larger than 1T}, Gemini-1.5~\cite{reid2024gemini}, and Llama3-8B-Instruct~\cite{llama3modelcard}. This baseline is set to compare the \method's ability to perform formal reasoning effectively.
(4) Mathematical LLMs: This baseline uses LLMs that are specifically trained in massive math-related corpus to perform whole proof generation. In particular, we choose: Llemma~\cite{azerbayev2023llemma}, and DeepSeek-Math-7B~\cite{shao2024deepseekmath} as the baselines. This aims to demonstrate under fine-tuned general-purpose LLMs can outperform math expert model under some cricumstances.

Following the results from~\cite{yang2024leandojo, polu2022formal}, we adopt pass@1 for all tree-search methods to ensure a consistent comparison under a similar GPU footprint during evaluation. The variation in the number of parameters for tree-search models also stems from our aim to maintain comparable computational costs. Specifically, our primary consideration for selecting baselines was to ensure the inference cost remains on the same scale, with an inference cost of 1.5 A6000 GPU days.

\subsection{Implementation Details}\label{exp:details}
The OBT dataset is generated using Gemini-1.5~\cite{reid2024gemini} for writing the natural language of the theorems and performing NL-FL bootstrapping. We use Gemnin-1.5 because it can give more human-like proofs and a better ability in NL-FL combination, details can be found in Appendix~\ref{appendix:deform}. The OBT dataset contains NL-FL aligned and bootstrapped 106,852 theorems, the data record format is in Appendix~\ref{appendix:record}. We perform Instruction Fine-tuning on Llama3-Instruct-8B~\cite{llama3modelcard} with 1,000 warm-up steps and a learning rate of 1E-5. Training takes approximately 32 hours on an 8 GPU A6000 machine. During the evaluation, we perform a uniform 128 generation for LLM's whole-proof generation. The initial examples for in-context learning are obtained from the proved theorem list of \citet{yang2024leandojo}. Depending on the context length, we use 10-16 examples for all LLM queries. 
We stop at the second round of iterative proof writing.

For few-shot LLM baselines, we use GPT-4-Turbo-0409 and Gemini-1.5-preview-0409 to perform formal reasoning. Since both models are released after Mathlib4, so they can have such data in their training set, which makes a fair comparison.

\subsection{Results}\label{exp:results}
We present the main experimental results in Tab.\ref{tab:main}. From the table, we can observe that \method achieves a cumulative accuracy rate of 36.48\% on MiniF2F-Valid and 33.61\% on MiniF2F-Test, suppressing all the baselines. 

It is also notable that the result of un-finetuned Llama3-8B-Instruct and GPT-4  have a similar accuracy rate on both the Test and Valid set despite the great difference in model size, this demonstrates most modern LLMs are under-trained on Lean4 reasoning. Surprisingly, Gemini has the best performance among all the baselines rather than GPT-4; this demonstrates its superior ability to understand formal language and gives indirect evidence that Gemini is a better choice of LLM to perform Informalization and Bootstrapping.

For the tree-search method, the large search space limits the choice of model in relatively small size and they only achieve an average 27.2\% accuracy rate, which is relatively low, demonstrating the limitation of such a method.

For the math-expert models, we can spot that \method also significantly outperforms them. This shows that large math-related code and text pertaining itself does not naturally provide the formal reasoning ability. This also demonstrates the effectiveness of our fine-tuning to transform a general-purpose model into a Lean4 expert.

\subsection{Ablation Studies}\label{exp:ablation}

\begin{table}
    \centering
    \resizebox{\linewidth}{!}
    {\begin{tabular}{ccc}
    \toprule
         \textbf{Method}             & \textbf{MiniF2F-Valid}    & \textbf{MiniF2F-Test}     \\
         \midrule
         \method                     & \textbf{34.84\%}         & \textbf{31.15\%}  \\
         w/o NL Guidance             & 24.18\%                  & 17.21\% \\
         w/o Bootstrapping           & 26.23\%                  & 26.23\% \\
         w/o Block Training          & 27.87\%                  & 23.36\% \\
         w/o Curriculum Data Sorting & 29.51\%                  & 25.83\% \\
    \bottomrule
    \end{tabular}}
    \caption{Ablation study result}
    \label{tab:abl}
    \vspace{-0.2in}
\end{table}

Due to the low GPU footprint of \method, we are able to perform a comprehensive ablation study to test the effectiveness of each component in the framework, namely, NL guidance, NL-FL bootstrapping, block training, and curriculum data sorting. 
In the ablation studies, we use the result of the first iteration with the default example list from~\citet{yang2024leandojo}.
The results are demonstrated in Tab.~\ref{tab:abl}. From the table, we can find that the removal of any component of our framework will lead to a significant performance drop compared to the full framework result, which proves that modules we propose are not trivial combinations, but work synergically to enhance the performance.

In the removal of NL Guidance, we perform the experiment under the setting without NL in training data but use the NL examples and NL guidance in the testing data. The accuracy rate dropped significantly, and the fine-tuned model does not outperform the untrained model. This indicates that merely more exposure to Lean4 proof code does not improve the ability of LLM in formal reasoning. When we remove the NL-FL bootstrapping, the performance drops because the LLM often loses track of the proof and keeps on generating the same tactic. With bootstrapping, the performance is much better due to the NL guidance.

The ablation study also shows that removing block training results in a performance drop, which we attribute to the distribution shift between training and testing data. Without block training, the training data lacks information about in-context examples, while the testing phase includes this in-context knowledge for the LLM to learn. Additionally, removing curriculum data sorting also leads to a performance decline. Curriculum data sorting provides a smoother training process by ensuring that lengthy and difficult examples do not appear early on and disrupt learning.

Despite performance drop when removing the individual component, with other components, our method still outperforms un-finetuned Llama3 except for without the NL guidance. It supports the effectiveness of other components from another perspective.

\subsection{Iterative Proof Writing study}\label{exp:iterative}

This section gives a closer look into the behavior of different rounds of iterative proof writing as demonstrated in the Tab.\ref{tab:iterative}. From the table, we can see that when using the example from the same dataset for proof writing, there is a 2\% accuracy increment in pass rate. This shows that our iterative proof writing method can reduce the distribution gap between testing data and examples to some extent.

\begin{table}
    \centering
    \resizebox{0.8 \linewidth}{!}
    {\begin{tabular}{ccc}
    \toprule
         \textbf{Method}             & \textbf{MiniF2F-Valid}    & \textbf{MiniF2F-Test}     \\
         \midrule
         Round 1                     & 34.84\%                   & 31.15\% \\
         Round 2                     & 36.48\%                   & 33.61\% \\
    \bottomrule
    \end{tabular}}
    \caption{Different rounds of iterative proof writing}
    \label{tab:iterative}
    \vspace{-0.2in}
\end{table}

\subsection{Effectiveness of Example Retrieval}\label{exp:T5}

This section studies the effectiveness of our fine-tuned T5 for example retrieval using graphical methods.   We encode all the formal theorem statements in Mathlib4 and the natural language theorem statement in the example list following Section~\ref{dataGen:deform}. We then compare the cosine similarity of all possible combinations of the formal and natural language theorem statement set. 

\begin{figure}
    \centering
    \resizebox{0.8\linewidth}{!}{\includegraphics{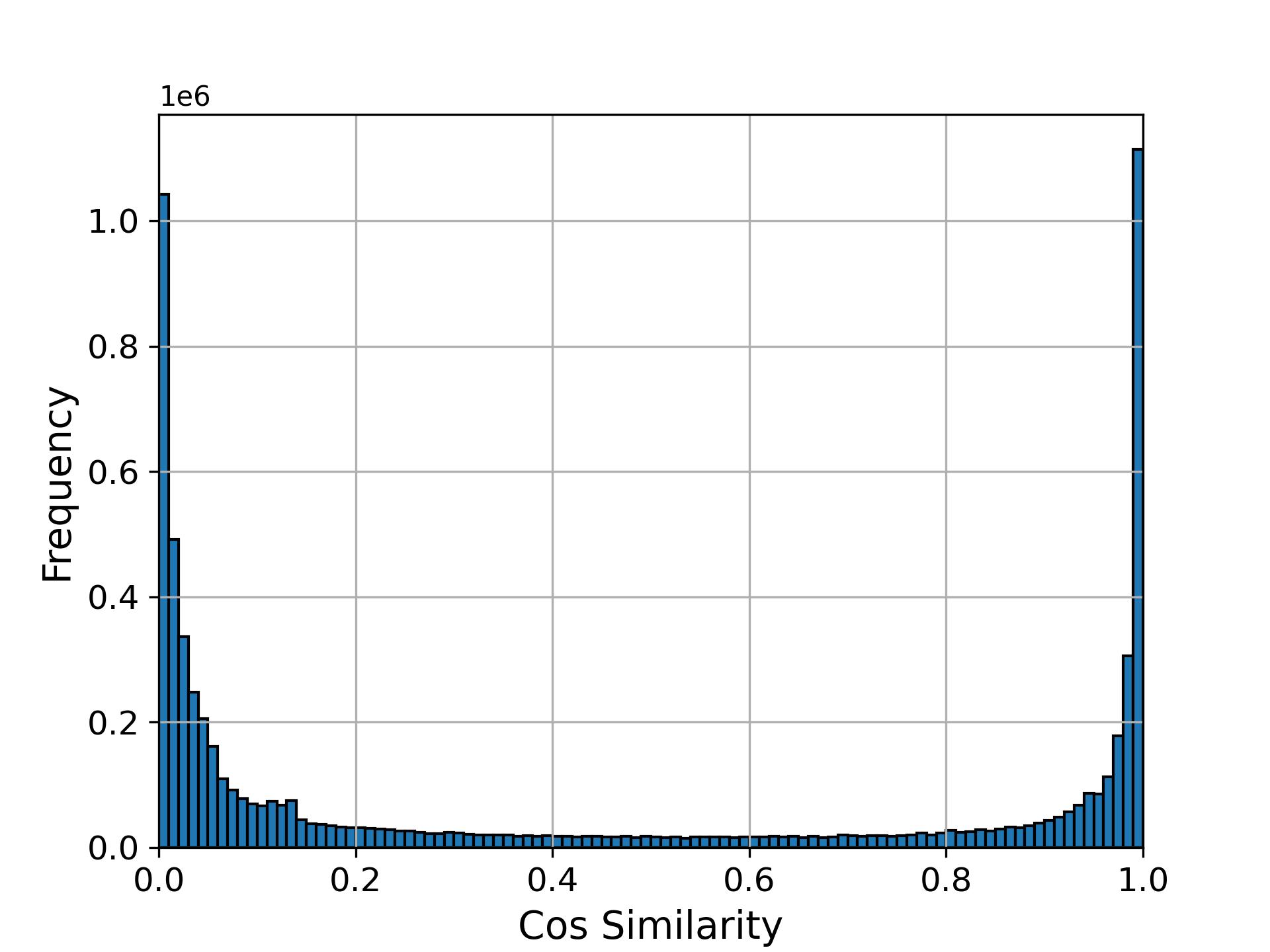}}
    \caption{Histogram for all combinations of NL statement in example list and FL statement in Mathlib4
    }
    \label{fig:histT5}
    \vspace{-0.2in}
\end{figure}

From the two peaks in the histogram at cos similarities of 1.0 and 0.0 in Fig.~\ref{fig:histT5}, we can see that our example selection model is certain about its prediction. Plus a low testing loss that we observed when training the model, means its performance is relatively good. We can say our retriever can largely  distinguish similar and different informal statements in the example list and formal statements in Mathlib4.

\subsection{Case study}\label{exp:case}

We analyze the behavior of \method by examining the proofs generated by the theorem prover from MiniF2F-Test in Lean4. Since MiniF2F-Test is a dataset that does not have publicly available proof, it is highly unlikely it will be included in the training dataset of any LLM. For more case studies, kindly refer to Appendix~\ref{appendix:case}.

\noindent\textbf{NL Statement:} Let $a$ and $b$ be two real numbers such that $a^2+b^2=1$. Show that $ab+(a-b) \leq 1$ 

\vspace{0.05in}

\noindent\textbf{NL Proof:} We have that $(a - b - 1)^2 \geq 0$. By expanding, we have: $0 \leq a^2 - ab - a - ab + b^2 + b - a + b + 1$. Since $a^2+b^2=1$, we get $0 \leq 2 - 2ab - 2a + 2b$. As a result, $ab+(a-b) \leq 1$.

\noindent\textbf{Generated Proof:}

\begin{lstlisting}
theorem algebra_sqineq_unitcircatbpamblt1
  (a b: ℝ)
  (h₀ : a^2 + b^2 = 1) :
  a * b + (a - b) ≤ 1 := by
  -- We have that (a - b - 1)^2 ≥ 0.
  have h₁ : 0 ≤ (a - b - 1) ^ 2 := sq_nonneg _
  -- By expanding, we have:
  -- 0 ≤ a^2 -ab-a-ab+b^2 +b-a+b+1.
  linarith [h₀, sub_add_cancel a b]
\end{lstlisting}
\vspace{-0.1in}

From the generated proof, we can see that the LLMs learn from the data to obtain the condition of $\geq 0$ in Lean and use the correct condition expression. Also, it learns to use the lean tactic \textit{linarith} to expand the inequality to obtain correct proof, which shows the Lean4 ability that LLM obtained in fine-tuning.
This example also demonstrates the effectiveness of using external and in-code NL proof to guide LLM in writing Lean4 proofs, which qualitatively validate the NL guidance and NL-FL bootstrapping method.

\section{Related work}\label{sec:relat}

\subsection{Formal Reasoning}\label{relat:formal}

Formal mathematical languages express mathematical statements in first-order logic. Such verification systems for mathematics are also known as \textit{Interactive Theorem Provers} (ITPs). There are many ITP languages such as Isabelle \cite{paulson1994isabelle}, Lean~\cite{de2015lean, moura2021lean}, Coq~\cite{coq1996coq}, Metamath~\cite{megill2019metamath}, and HOL Light \cite{harrison2009hol}. The formal languages embed mathematical definitions and theorems onto a concrete logical foundation in their kernels. Following \citet{polu2022formal}, we work on Lean because Lean proves are typically more concise and relatively understudied.

Many works focus on automatically completing formal reasoning using machine learning-based methods. Some use more traditional methods like K Nearest Neighbor (KNN)~\cite{gauthier2021tactictoe} or Graph Neural Network (GNN)~\cite{yang2019learning}. Others take advantage of the recent development of deep transformer-based methods that treat theorems as plain texts. Among them, Expert Iteration~\cite{polu2022formal} and ReProver~\cite{yang2024leandojo} focus on training existing LLMs to generate tactics and perform a tree search to complete the proofs. Other methods focus on exploring the few-shot capability, allowing LLMs to directly generate the whole proof based on the guidance of natural language \cite{wu2022autoformalization, jiang2022draft}. Although there are some attempts to construct aligned datasets for theorem proofing~\cite{azerbayev2023proofnet, ying2024lean}, there is still no large-scale NL-FL aligned dataset. This leads to the fact that there are currently no universally recognized methods for training LLMs to generate whole proof directly.

\subsection{Dataset Generation}\label{relat:dataset}
Modern machine learning methods typically require massive datasets to learn an application.
However, it is impractical to have high-quality data for every corner case, prompting researchers to explore dataset generation.
By combining existing incomplete data with the rich knowledge in LLMs, dataset generation can leverage this knowledge to produce a complete dataset suitable for model training. Initial attempts have achieved this through fine-tuned generative models~\cite{anaby2020not, chen2020big}. Other researchers explore zero-shot or few-shot performance for modern LLMs by directly querying the LLMs to obtain the intended dataset~\cite{meng2022generating, gao2022self, wang2023let}. 
In this work, we take advantage of these ideas for dataset generation to obtain the OBT dataset.

\section{Conclusion}\label{sec:conc}

This paper proposes \method, an end-to-end framework for transforming a general-purpose LLM into a Lean4 expert, along with the \textit{Open Bootstrapped Theorems} (OBT) dataset, a NL-FL aligned, bootstrapped dataset for training an LLM Lean4 prover. This work largely addresses the significant data scarcity problem by introducing the NL-FL Aligned Dataset Generation method, which is used to create the OBT dataset. Subsequently, we demonstrate block training and curriculum data sorting techniques to enhance LLM training. Furthermore, we present the Iterative Proof Writing tactic to better utilize the LLM's capability in theorem proving. The major innovation of this work is the NL-FL bootstrapping method, which enables the LLM to better transfer its natural language reasoning ability to Lean4 proof writing through generated data. We also conduct comprehensive experiments to evaluate the effectiveness of \method, where our framework successfully proves 36.48\% and 33.61\% of the theorems in MiniF2F-Valid and Test, respectively, surpassing the GPT-4 and Gemini-1.5 baselines. We will open-source all the datasets to facilitate further development in the field.

\section{Discussion}
Although large-scale pre-train provides LLMs with strong abilities in most general tasks, there are many corner cases that lack data for any machine learning methods to be effective. Formal reasoning is one of the most significant examples. From a border perspective, \method sheds light on a general framework to further apply modern LLMs to such corner cases. It contains methods to leverage existing incomplete data, techniques to better train LLMs for unfamiliar tasks, and strategies to enhance LLM's performance in application. Thus, the contribution of this paper is not limited to the field of formal reasoning but gives a general framework for the further usage of LLMs in corner cases.

\section*{Acknowledgement}
We thank the anonymous reviewers for their feedback on our paper. We also want to express our greatest gratitude to Professor Frederick Tsz-Ho Fong at HKUST for his generous support in the mathematical aspect of this paper. 

\section*{Limitations}
Despite the promising results of \method, there are still some limitations in our work that can be addressed in future research. First, even with natural language proofs to guide Lean4 proof writing, all existing formal provers, including \method, struggle with solving difficult IMO-level problems. We conclude that LLMs currently lack the ability to understand the intricate technical aspects of human proofs. Integrating the "kindles" in human-written proofs into LLMs is an overlooked area in current research. Secondly, due to the complexity of the Lean4 kernel, this paper does not explore the potential of RL methods for enabling LLMs to write formal proofs through online interaction with Lean, nor does it incorporate feedback from Lean to refine incorrect proofs. This requires deeper integration of Lean and Python. Thirdly, although formal language provides a concrete foundation for verifying the correctness of mathematical proofs, there are potential risks that a natural language-guided Lean4 prover may automatically correct some errors in natural language. This could lead to errors in natural language being considered correct and cause wrong natural language proofs to be subsequently propagated within the mathematical community.

\bibliography{custom}

\newpage
\appendix\label{sec:appendix}

\section{Confusing Lean3 Data in LLM Pre-train}\label{appendix:pretrain}

While studying how to directly generate Lean4 formal proofs using LLMs, we found that most LLMs have a serious problem with hallucination. The most significant issue is that, even with clear prompt instructions, the LLMs tend to write Lean3 proofs that are incompatible with Lean4.

The results for GPT-4 are shown in Tab.~\ref{tab:pretrain:gpt4-1} and \ref{tab:pretrain:gpt4-2}. From these tables, we can observe that even with clear instructions to use Lean4 for writing proofs, the LLM still uses Lean3 syntax for all imports and proofs. The imports are from Lean3 repositories or sometimes do not exist, and the proof segments, indicated by "begin" and "end," are also from Lean3. This issue also occurs with Llama-3-8B-Instruct and Gemini-1.5-Pro, but less frequently. We attribute this behavior to the massive amount of Lean3 data used in the pre-training of these LLMs. This causes LLMs to fail in fully utilizing their knowledge in formal reasoning, as many generated formal proofs are incorrect in format.

Alternatively, \method uses extensive Lean4 data during instruction fine-tuning to significantly reduce this problem, as detailed in Section~\ref{exp:case} and Appendix~\ref{appendix:case}.

\section{Case Study}\label{appendix:case}

This section provides additional case studies to further evaluate the performance of \method in proving Lean4 theorems with NL guidance. Here, we select most examples from MiniF2F-Valid to avoid revealing too much proof information about MiniF2F-Test and contaminating the dataset. We present the examples in Tab.~\ref{tab:case-1}, \ref{tab:case-2}, \ref{tab:case-3}, \ref{tab:case-4}, and \ref{tab:case-5}.

From case 1 in Tab.~\ref{tab:case-1}, we can see that the LLMs learn how to perform complex reductions stated in Lean4 code based on the natural language. The "calc" section demonstrates the LLM's ability to correctly reduce algebraic formulas based on conditions that are not explicitly stated in the natural language proof.

Case 2 in Tab.~\ref{tab:case-2} demonstrates that under the \method framework, the LLM learns how to add sub-goals for proving in the correct Lean4 form from natural language proof. This is not observed in any of the correct proofs in the untrained model.

From cases 3, 4, and 5 in Tab.~\ref{tab:case-3}, \ref{tab:case-4}, and \ref{tab:case-5}, we can see the ability of our LLM to perform step-by-step reasoning in both natural language and formal language in relatively complex proofs, demonstrating the effectiveness of NL-FL bootstrapping.

\section{Various dataset size training}

This section presents the results of training the model on different scales of the OBT dataset. Specifically, we randomly sampled 1k and 10k subsets from the OBT dataset, and the results are shown in Tab.~\ref{tab:size}.

From the table, we observe that the performance improvement does not align with typical human behavior in text writing, resulting in the performance increase not following the expected scaling law. This phenomenon is also observed in dataset generation works such as \citet{wang2023let}.

\begin{table}
    \centering
    \resizebox{0.9 \linewidth}{!}
    {\begin{tabular}{cccc}
    \toprule
                                     & \textbf{1k}    & \textbf{10k}    & \textbf{full OBT (~100k)}     \\
         \midrule
         MiniF2F-Test                & 29.01\%        & 30.74\%         & 31.15\% \\
    \bottomrule
    \end{tabular}}
    \caption{Various dataset size}
    \label{tab:size}
\end{table}

\section{Different LLM's behavior in Informalization}\label{appendix:deform}

This section details why we use Gemini-1.5 as the LLM for informalization and NL-FL cootstrapping rather than the commonly used OpenAI GPT family models through examples. We demonstrate an example of informalization in Table~\ref{tab:deform}. From the table, we can see that the GPT-4-generated proof is more like an explanation of the Lean4 code, while the Gemini-1.5-generated proof resembles the wording commonly used in mathematics. This demonstrates that Gemini-1.5 has a better ability to understand Lean4 code for informalization. This is also indirectly supported by Gemini's superior performance in writing Lean4 proofs for the MiniF2F dataset, as shown in Table~\ref{tab:main}.

\section{OBT dataset record}\label{appendix:record}

The data record in OBT contains the following components:
\begin{enumerate}
    \item \textbf{Name:} the name of the theorem, following the name of dataset extracted in~\citet{yang2024leandojo}
    \item \textbf{Statement:} Lean4 statement of the theorem
    \item \textbf{Proof:} Lean4 theorem statement together with the proof, directly extracted from Mathlib4
    \item \textbf{File\_path:} The git repository for the given data record (for OBT dataset, it should be Mathlib4)
    \item \textbf{Commit:} The exact commit number of the theorem
    \item \textbf{Generated\_informal\_statement\_and\_proof:} The generated theorem informal theorem statement and proof
    \item \textbf{Commented\_proof:} The NL-FL bootstrapped Lean4 code.
\end{enumerate}
the example of an OBT dtaset record is presented at Tab.~\ref{tab:OBTexample}

\section{More bootstrapping method analysis}

This section studies the differences between various data bootstrapping methods. Our NL-FL Bootstrapping method is inspired by in-code comments, which can help the LLM transfer its ability from natural language to formal language. To demonstrate the effectiveness of this approach, we experimented with another bootstrapping method, namely Head Bootstrapping. This method involves adding natural language reasoning as comments before the formal language in the training data, allowing the model to learn this behavior and potentially encouraging deeper reasoning. We tested this approach using blocked training and a curriculum-learning strategy. The Head Bootstrapping method achieved a result of 25.46\%, while our NL-FL Bootstrapping method reached 31.15\%. This comparison demonstrates the effectiveness of our bootstrapping method over other dataset bootstrapping techniques.

\section{Applying TheoremLlama framework to other models}
To evaluate the generality of the TheoremLlama framework, we apply the dataset generated, together with curriculum learning technique and block training method to fine-tune other models. In particular, we choose DeepSeek-Math-7B~\cite{shao2024deepseekmath} as the base model for training, the result demonstrated in Tab.\ref{tab:deepseek}.

From the table, we can see that after applying our method to the math expert, the Lean4 performance of the model has been greatly increased. This proves \method can be effective in a wide range of models rather than could be only applied to Llama.

\begin{table}
    \centering
    \resizebox{\linewidth}{!}
    {\begin{tabular}{ccc}
    \toprule
         \textbf{Method}                & \textbf{MiniF2F-Valid}    & \textbf{MiniF2F-Test}     \\
         \midrule
         DeepSeek-Math-7B               & 25.80\%                   & 24.60\% \\
         TheoremLlama + DeepSeek-Math   & 38.52\%                   & 35.66\% \\
    \bottomrule
    \end{tabular}}
    \caption{Applying TheoremLlama method to DeepSeek-Math}
    \label{tab:deepseek}
    \vspace{-0.2in}
\end{table}

\section{Analyze the impact of NL-FL Bootstrapping}
This section provides a example towards the NL-FL Bootstrapping. 

Generated theorem proof without Bootstrapping:

\begin{lstlisting}
theorem amc12b_2002_p2
  (x : ℤ)
  (h₀ : x = 4) :
  (3 * x - 2) * (4 * x + 1) - (3 * x - 2) * (4 * x) + 1 = 11 := by
  subst x
  ring
\end{lstlisting}

Generated theorem proof with Bootstrapping:

\begin{lstlisting}
theorem algebra_sqineq_unitcircatbpamblt1
  (a b: ℝ)
  (h₀ : a^2 + b^2 = 1) :
  a * b + (a - b) ≤ 1 := by
  -- We have that (a - b - 1)^2 ≥ 0.
  have h₁ : 0 ≤ (a - b - 1) ^ 2 := sq_nonneg _
  -- By expanding, we have:
  -- 0 ≤ a^2 -ab-a-ab+b^2 +b-a+b+1.
  linarith [h₀, sub_add_cancel a b]
\end{lstlisting}

We can clearly find out that the generated result without bootstrapping tends not to use natural language to guide the generation of proof, which makes it fail to solve the problem algebra\_sqineq\_unitcircatbpamblt1. Thus, both ablation studies in Section~\ref{tab:abl} and case studies can provide the effectiveness of the bootstrapping method.

\section{OBT dataset quality analysis}\label{appendix:OBTQuality}


This section analyzes the quality of the OBT dataset. Due to the large volume of data, it is infeasible to analyze every record to verify if it has been informalized correctly. Therefore, we randomly sampled 40 records and evaluated them manually. Among these, 29 are correct, 4 are partially correct, and 7 contain critical mistakes that render the proof mathematically incorrect. The major mistake the LLM makes is confusing some complex mathematical concepts, such as mixing the general Lebesgue measure with the probability measure. There are also misunderstandings in vital parts of formal reasoning, for instance, mistakenly using normal sequence convergence in natural language reasoning while the formal language uses the Cauchy sequence to prove sequence convergence.

\section{Term List}\label{appendix:term}

Since formal theorem proving is a relatively new field, there is many fundamental terms that are hard to remember. Thus, we provide a table explaining some important terms in this paper that are not fully elaborated in Tab.\ref{tab:termList} for researchers who are not that familiar with that field to read our paper smoother.

\begin{table*}
    \centering
    \begin{tabular}{p{1.5cm} p{12cm}}
        \toprule
        \textbf{Prompt}             & You are a Lean4 expert who can write good Lean4 code based on natural language mathematical theorem and proof \\
                                    & <Examples> \\
                                    & \#\#\# Natural language version of theorem and proof: <NL of amc12a\_2019\_p21> \\
                                    & \#\#\# Lean4 version of theorem statement: <Lean4 statement of amc12a\_2019\_p21> \\
                                    & 
\begin{lstlisting}
theorem amc12a_2019_p21 (z : ℂ) (h₀ : z = (1 + Complex.I) / Real.sqrt 2) :
  ((∑ k : ℤ in Finset.Icc 1 12, z ^ k ^ 2) * (∑ k : ℤ in Finset.Icc 1 12, 1 / z ^ k ^ 2)) = 36 :=
\end{lstlisting}\\
                                    & \#\#\# Lean4 version of theorem and proof: \\
        \midrule
        \textbf{Generated Result}   & To translate the given natural language explanation of the theorem into a Lean4 proof, we must first define the necessary elements accurately in Lean4 syntax. This includes the expression for $z$, the computation of powers and their modulo reductions, and ultimately the statement of the sums and their product. Lean4 provides a robust framework for handling complex numbers and integer arithmetic, which we will use here. \\
                                    & 
\begin{lstlisting}
import data.complex.basic
import data.finset.basic
import data.nat.basic

open complex

-- Define the main theorem
theorem amc12a_2019_p21 : 
  let z := (1 + I) / sqrt 2 in
  let s := ∑ k in finset.range 12, (z ^ (k+1) ^ 2) in
  let t := ∑ k in finset.range 12, (1 / (z ^ (k+1) ^ 2)) in
  s * t = 36 :=
begin
  ... -- Details omitted
end
\end{lstlisting} \\
    \bottomrule
    \end{tabular}
    \caption{GPT-4 writing Lean4 proof Example 1}
    \label{tab:pretrain:gpt4-1}
\end{table*}

\begin{table*}
    \centering
    \begin{tabular}{p{1.5cm} p{12cm}}
    \toprule
    \textbf{Prompt}     & You are a Lean4 expert who can write good Lean4 code based on natural language mathematical theorem and proof \\
                        & <Examples> \\
                        & Natural language version of theorem and proof: \\
                        & mathd\_numbertheory\_543: Find the number of distinct positive divisors of $(30)^4$ excluding 1 and $(30)^4$. Show that it is 123. \\
                        & $$ (30^4) = (2^1 \cdot 3^1 \cdot 5^1)^4 = 2^4 \cdot 3^4 \cdot 5^4 $$Since $t(30^4) = (4+1)^3 = 125$, taking out 1 and $(30^4)$ leaves $125 - 2 = 123$ positive divisors. \\
                        & \#\#\# Lean4 version of theorem statement: \\
                        & 
\begin{lstlisting}
    theorem mathd_numbertheory_543 : (∑ k in Nat.divisors (30 ^ 4), 1) - 2 = 123 :=
\end{lstlisting} \\

                                    & \#\#\# Lean4 version of theorem and proof: \\
    \midrule
    \textbf{Generated Result}       & To formalize and prove the given theorem in Lean4, we need to use Lean's mathematical libraries, particularly those related to natural numbers and their properties, such as divisor functions. The theorem involves computing the total number of divisors of $30^4$, excluding 1 and $30^4$ itself. \\
                                    & 
\begin{lstlisting}
import data.nat.prime
import algebra.big_operators

open_locale big_operators

-- Define the theorem stating that the number of divisors of 30^4, excluding 1 and 30^4, is 123.
theorem mathd_numbertheory_543 : 
  (∑ k in nat.divisors (30^4), 1) - 2 = 123 :=
begin
    ... -- Details omitted
end
\end{lstlisting} \\
    \bottomrule
    \end{tabular}
    \caption{GPT-4 writing Lean4 proof Example 2}
    \label{tab:pretrain:gpt4-2}
\end{table*}

\begin{table*}
    \centering
    \begin{tabular}{p{3cm} p{12cm}}
    \toprule
    \textbf{NL Statement}           & If $f(x) = \dfrac{1}{x + 2},$ what is $f(f(1))$? Show that it is $\dfrac{3}{7}$. \\
    \midrule
    \textbf{NL Proof}               & We see that $f(1) = \dfrac{1}{1 + 2} = \dfrac{1}{3}.$ Therefore, $f(f(1)) = f\left(\dfrac{1}{3}\right) = \dfrac{1}{\frac{1}{3} + 2} = \dfrac{1}{\frac{7}{3}} = \dfrac{3}{7}.$ \\
    \midrule
    \textbf{Generated Proof}        & 
\begin{lstlisting}
theorem mathd_algebra_270
  (f : ℝ → ℝ)
  (h₀ : ∀ x, x ≠ -2 -> f x = 1 / (x + 2)) :
  f (f 1) = 3/7 := by
  -- We see that f 1 = 1 / (1 + 2) = 1 / 3
  have h₁ : f 1 = 1 / 3 := by norm_num [h₀]
  -- Thus f (f 1) = f (1 / 3) = 1 / (1 / 3 + 2) = 1 / (7 / 3) = 3 / 7
  calc
    f (f 1) = f (1 / 3) := by rw [h₁]
    _ = 1 / (1 / 3 + 2) := by norm_num [h₀]
    _ = 1 / (7 / 3) := by norm_num
    _ = 3 / 7 := by norm_num
\end{lstlisting} \\
    \bottomrule
    \end{tabular}
    \caption{Additional Case Study 1}
    \label{tab:case-1}
\end{table*}

\begin{table*}
    \centering
    \begin{tabular}{p{3cm} p{12cm}}
    \toprule
    \textbf{NL Statement}           & If $|x - 2| = p$, where $x < 2$, then $x - p = $ \\
                                    & $\textbf{(A)} -2 \qquad \textbf{(B)} 2 \qquad \textbf{(C)} 2-2p \qquad \textbf{(D)}  2p-2 \qquad \textbf{(E)} |2p-2|  $ \\
                                    & Show that it is $\textbf{(C)} 2-2p$.\\
    \midrule
    \textbf{NL Proof}               & When $x < 2,$ $x-2$ is negative so $|x - 2| = 2-x = p$ and $x = 2-p$. Thus $x-p = (2-p)-p = 2-2p$. Leading to answer $\textbf{(C)} 2-2p$ \\
    \midrule
    \textbf{Generated Proof}        & 
\begin{lstlisting}
theorem amc12_2000_p5 (x p : ℝ) (h₀ : x < 2) (h₁ : abs (x - 2) = p) : x - p = 2 - 2 * p := by
  -- If x < 2, then x - 2 is negative, so |x - 2| = 2 - x = p.
  -- Thus, x = 2 - p.
  suffices abs (x - 2) = -(x - 2) by
    rw [h₁] at this
    linarith
  -- Therefore, x - p = (2 - p) - p = 2 - 2p.
  apply abs_of_neg
  linarith
\end{lstlisting} \\
    \bottomrule
    \end{tabular}
    \caption{Additional Case Study 2}
    \label{tab:case-2}
\end{table*}

\begin{table*}
    \centering
    \begin{tabular}{p{3cm} p{12cm}}
    \toprule
    \textbf{NL Statement}           & Suppose that $g(x)=f^{-1}(x)$. If $g(-15)=0$, $g(0)=3$, $g(3)=9$ and $g(9)=20$, what is $f(f(9))$? Show that it is 0. \\
    \midrule
    \textbf{NL Proof}               & Since $f$ and $g$ are inverses and $g(3) = 9$, we have $f(9) = 3$, so $f(f(9)) = f(3)$. Similarly, $g(0) = 3$, so $f(3) = 0$. \\
    \midrule
    \textbf{Generated Proof}        & 
\begin{lstlisting}
theorem mathd_algebra_451 
    (σ : Equiv ℝ ℝ) 
    (h₀ : σ.2 (-15) = 0) 
    (h₁ : σ.2 0 = 3) 
    (h₂ : σ.2 3 = 9)
    (h₃ : σ.2 9 = 20) : σ.1 (σ.1 9) = 0 := by
  -- Since f and g are inverses and g(3) = 9, we have f(9) = 3, so f(f(9)) = f(3).
  simp only [Equiv.invFun_as_coe, eq_comm] at h₀ h₁ h₂ h₃
  -- Similarly, g(0) = 3, so f(3) = 0.
  simp only [Equiv.toFun_as_coe]
  rw [← Equiv.apply_eq_iff_eq_symm_apply σ] at h₂
  rw [← Equiv.apply_eq_iff_eq_symm_apply σ] at h₁
  have h₄ := (Equiv.apply_eq_iff_eq σ).mpr h₂
  rw [h₁] at h₄
  exact h₄
\end{lstlisting} \\
    \bottomrule
    \end{tabular}
    \caption{Additional Case Study 3}
    \label{tab:case-3}
\end{table*}

\begin{table*}
    \centering
    \begin{tabular}{p{3cm} p{12cm}}
    \toprule
    \textbf{NL Statement}           & For what real value of $k$ is $\frac{13-\sqrt{131}}{4}$ a root of $2x^2-13x+k$? Show that it is $\frac{19}{4}$. \\
    \midrule
    \textbf{NL Proof}               & We could substitute $(13-\sqrt{131})/4$ for $x$ in the equation, but the quadratic formula suggests a quicker approach. Substituting $2$, $-13$, and $k$ into the quadratic formula gives  
    $$\frac{-(-13)\pm\sqrt{(-13)^2-4(2)(k)}}{2(2)}= \frac{13\pm\sqrt{169-8k}}{4}.$$
    Setting $(13+\sqrt{169-8k})/4$ and $(13-\sqrt{169-8k})/4$ equal to $(13-\sqrt{131})/4$, we find no solution in the first case and $169-8k=131$ in the second case.  Solving yields $k=(169-131)/8=38/8=\frac{19}{4}$. \\
    \midrule
    \textbf{Generated Proof}        & 
\begin{lstlisting}
theorem mathd_algebra_116 (k x : ℝ) (h₀ : x = (13 - Real.sqrt 131) / 4)
    (h₁ : 2 * x ^ 2 - 13 * x + k = 0) : k = 19 / 4 := by
  -- Proof: We are given that (13 - $\sqrt{131}$) / 4 is a root of the quadratic 2x² - 13x + k = 0
  -- and want to show that k = 19/4.
  rw [h₀] at h₁ -- Substitute (13 - $\sqrt{131}$) / 4 for x in the quadratic equation.
  -- We now have a equation that is reducible to k = 19/4.
  rw [eq_comm.mp (add_eq_zero_iff_neg_eq.mp h₁)] -- Rearrange the equation obtained from the previous step.
  norm_num -- Normalize the numeric expressions.
  rw [pow_two] -- Expand the square term.
  rw [mul_sub] -- Expand by distributivity.
  rw [sub_mul, sub_mul] -- Expand by distributivity.
  rw [Real.mul_self_sqrt _] -- Simplify $\sqrt{a}$ * $\sqrt{a}$ to a.
  ring -- Apply the ring axioms to simplify the expression.
  linarith -- Verify that the left and right sides of the equation are equal, thus proving k = 19/4
\end{lstlisting} \\
    \bottomrule
    \end{tabular}
    \caption{Additional Case Study 4}
    \label{tab:case-4}
\end{table*}

\begin{table*}
    \centering
    \begin{tabular}{p{3cm} p{12cm}}
    \toprule
    \textbf{NL Statement}           & If $3a + b + c = -3, a + 3b + c = 9, a + b + 3c = 19$, then find $abc$. Show that it is -56. \\
    \midrule
    \textbf{NL Proof}               & Summing all three equations yields that $5a + 5b + 5c = -3 + 9 + 19 = 25$. Thus, $a + b + c = 5$. Subtracting this from each of the given equations, we obtain that $2a = -8, 2b = 4, 2c = 14$. Thus, $a = -4, b = 2, c =7$, and their product is $abc = -4 \times 2 \times 7 = -56$. \\
    \midrule
    \textbf{Generated Proof}        & 
\begin{lstlisting}
theorem mathd_algebra_338
  (a b c : ℝ)
  (h₀ : 3 * a + b + c = -3)
  (h₁ : a + 3 * b + c = 9)
  (h₂ : a + b + 3 * c = 19) :
  a * b * c = -56 := by
  -- This theorem shows that if 3a + b + c = -3, a+3b+c = 9, a+b+3c = 19,
  -- then a * b * c = -56.
  have h₃ : a + b + c = 5 := by linarith
  -- From the first equation, 3a + b + c = -3, we have a + b + c = 5.
  have h₄ : 2 * a = -8 := by linarith
  -- From the first equation, 3a + b + c = -3, we also have 2 * a = -8.
  have h₅ : 2 * b = 4 := by linarith
  -- From the second equation, a+3b+c = 9, we have 2 * b = 4.
  have h₆ : 2 * c = 14 := by linarith
  -- From the third equation, a+b+3c = 19, we have 2 * c = 14.
  have h₇ : a = -4 := by linarith
  -- From h₄ and h₃, we have a = -4.
  have h₈ : b = 2 := by linarith
  -- From h₅ and h₃, we have b = 2.
  have h₉ : c = 7 := by linarith
  -- From h₆ and h₃, we have c = 7.
  simp_all only [mul_neg, neg_mul, mul_assoc, neg_add, add_assoc, add_left_comm, sub_eq_add_neg,
    sub_neg_eq_add, eq_self_iff_true, true_and]
  ring_nf
  -- Finally, we can conclude that a * b * c = -4 * 2 * 7 = -56.
  -- QED.

\end{lstlisting} \\
    \bottomrule
    \end{tabular}
    \caption{Additional Case Study 5}
    \label{tab:case-5}
\end{table*}

\begin{table*}
    \centering
    \hspace*{-1cm}
    \begin{tabular}{p{1.5cm} p{12cm}}
    \toprule
    \textbf{Prompt}         & You are a mathematician who can write natural language proof based on Lean4 proof \\
                            & <Examples> \\
                            & Lean4 version of theorem statement and proof: \\
                            & 
\begin{lstlisting}
theorem integral_eq_sub_of_hasDerivAt (hderiv : ∀ x ∈ uIcc a b, HasDerivAt f (f' x) x)
    (hint : IntervalIntegrable f' volume a b) : ∫ y in a..b, f' y = f b - f a :=
  integral_eq_sub_of_hasDeriv_right (HasDerivAt.continuousOn hderiv)
    (fun _x hx => (hderiv _ (mem_Icc_of_Ioo hx)).hasDerivWithinAt) hint
\end{lstlisting} \\
                            & Natural language version of theorem statement and proof: \\
    \midrule
    \textbf{Gemini Result}  & \textbf{Statement:} Let \( f : \mathbb{R} \to \mathbb{R} \) be a function that is differentiable on the interval \([a, b]\). We want to show that $\int_a^b f'(x) \, dx = f(b) - f(a).$ \\
                            & \textbf{Proof:} Since $f$ is differentiable on \([a, b]\), it is continuous on \([a, b]\). By the Fundamental Theorem of Calculus, Part 1, we know that if \( F \) is an antiderivative of \( f' \) on \([a, b]\), then 
                            $\int_a^x f'(t) \, dt = F(x) - F(a)$
                            for any \( x \in [a, b] \). In particular, if we take \( x = b \), we get
                            $\int_a^b f'(t) \, dt = F(b) - F(a).$
                            Since \( F \) is an antiderivative of \( f' \), we can choose \( F \) to be \( f \) itself. Therefore, we have
                            $F(x) = f(x)$
                            and the equation becomes
                            $\int_a^b f'(t) \, dt = f(b) - f(a). $
                            Thus, we have shown that
                            $ \int_a^b f'(x) \, dx = f(b) - f(a), $
                            which completes the proof. \\
    \midrule
    \textbf{GPT Result}     & \textbf{Statement:} Given a function \( f \) defined on an interval \([a, b]\), and \( f' \), its derivative, prove that the integral of \( f' \) from \( a \) to \( b \) equals \( f(b) - f(a) \). \\
                            & \textbf{Proof:} Assume that for every point \( x \) in the closed interval \([a, b]\), the function \( f \) has a derivative \( f'(x) \) at \( x \), denoted by \( \text{HasDerivAt } f (f'(x)) x \). Also, assume that \( f' \) is integrable over the interval from \( a \) to \( b \), denoted by \( \text{IntervalIntegrable } f' \, \text{volume} \, a \, b \). \\
                            & To prove the theorem, we utilize the Fundamental Theorem of Calculus, which states that if a function \( g \) is continuous on \([a, b]\) and has an antiderivative \( F \) over that interval, then 
                            $\int_a^b g(x) \, dx = F(b) - F(a).$
                            In this case, \( f' \) acts as \( g \), and its antiderivative is \( f \) by the definition of a derivative. The condition \( \text{HasDerivAt } f (f'(x)) x \) implies that \( f' \) is continuous on \([a, b]\) due to the continuous nature of derivatives under normal circumstances. \\
                            & \textbf{1. Continuity of \( f' \):} The hypothesis \( \forall x \in [a, b], \text{HasDerivAt } f (f'(x)) x \) assures that \( f' \) is continuous on \([a, b]\). This is because the derivative at a point implies local continuity at that point. \\
                            & \textbf{2. Integral of \( f' \):} Since \( f' \) is continuous over \([a, b]\) and integrable (as given by \( \text{IntervalIntegrable } f' \, \text{volume} \, a \, b \)), we apply the Fundamental Theorem of Calculus. Hence,
                            $\int_a^b f'(y) \, dy = f(b) - f(a). $ \\
                            & This concludes the proof. The integral of the derivative \( f' \) from \( a \) to \( b \) results in the net change in the function \( f \) over the interval, which is \( f(b) - f(a) \). \\
    \bottomrule
    \end{tabular}
    \caption{Comparison of Gemini-1.5 and GPT-4 generated NL proofs}
    \label{tab:deform}
\end{table*}

\begin{table*}
    \centering
    \hspace*{-1cm}
    \begin{tabular}{p{1.5cm} p{12cm}}
    \toprule
    \textbf{Name}                    & intervalIntegral.integral\_eq\_sub\_of\_hasDerivAt \\
    \midrule
    \textbf{Statement}               & 
\begin{lstlisting}
theorem integral_eq_sub_of_hasDerivAt (hderiv : ∀ x ∈ uIcc a b, HasDerivAt f (f' x) x)
    (hint : IntervalIntegrable f' volume a b) : ∫ y in a..b, f' y = f b - f a :=
\end{lstlisting} \\
    \midrule
    \textbf{Proof}                   & 
\begin{lstlisting}
theorem integral_eq_sub_of_hasDerivAt (hderiv : ∀ x ∈ uIcc a b, HasDerivAt f (f' x) x)
    (hint : IntervalIntegrable f' volume a b) : ∫ y in a..b, f' y = f b - f a :=
    integral_eq_sub_of_hasDeriv_right (HasDerivAt.continuousOn hderiv)
    (fun _x hx => (hderiv _ (mem_Icc_of_Ioo hx)).hasDerivWithinAt) hint
\end{lstlisting} \\
    \midrule
    \textbf{File\_path}             & https://github.com/leanprover-community/mathlib4 \\
    \midrule
    \textbf{Commit}                 & 3ce43c18f614b76e161f911b75a3e1ef641620ff \\
    \midrule
    \textbf{Generated\_ informal\_ statement\_ and\_ proof} 
        & Let $f : \mathbb{R} \to \mathbb{R}$ be a function that is differentiable on the interval $[a, b]$. Use the Fundamental Theorem of Calculus to show that$$\int_a^b f'(x) dx = f(b) - f(a).$$ \\
        & \textbf{Proof:} Since $f$ is differentiable on $[a, b]$, it is continuous on $[a, b]$. By the Second Fundamental Theorem of Calculus, we have
        $\frac{d}{dx} \int_a^x f'(t) dt = f'(x) \quad \forall x \in [a, b].$
        Integrating both sides of this equation with respect to $x$ from $a$ to $b$, we get
        $\int_a^b \frac{d}{dx} \int_a^x f'(t) dt dx = \int_a^b f'(x) dx.$
        But the left-hand side of this equation is equal to $\int_a^b f'(x) dx$, so we have
        $\int_a^b f'(x) dx = \int_a^b f'(x) dx.$
        Subtracting $\int_a^b f'(x) dx$ from both sides of this equation, we get
        $0 = f(b) - f(a).$
        Therefore, $f(b) - f(a) = 0$, which proves the desired result. \\
    \midrule
    \textbf{Commented\_proof:}      & 
\begin{lstlisting}
theorem integral_eq_sub_of_hasDerivAt (hderiv : ∀ x ∈ uIcc a b, HasDerivAt f (f' x) x)
    (hint : IntervalIntegrable f' volume a b) : ∫ y in a..b, f' y = f b - f a :=
    -- This theorem states that the integral of the derivative of a function over an interval
    -- is equal to the difference of the function values at the endpoints of the interval.
    -- This is a fundamental theorem of calculus.
    integral_eq_sub_of_hasDeriv_right (HasDerivAt.continuousOn hderiv)
    -- This part of the proof uses the second fundamental theorem of calculus, which states that
    -- the derivative of the integral of a function is equal to the function itself.
    (fun _x hx => (hderiv _ (mem_Icc_of_Ioo hx)).hasDerivWithinAt) hint
    -- This part of the proof uses the fact that the derivative of a function is continuous on
    -- the interval where it is differentiable.
\end{lstlisting} \\
    \bottomrule
    \end{tabular}
    \caption{Example of one record of OBT dataset}
    \label{tab:OBTexample}
\end{table*}

\begin{table*}
    \centering
    \begin{tabular}{p{3cm} p{9.5cm}}
        \toprule
        \textbf{Term}           & \textbf{Explaination} \\
        \midrule
        Lean4 \& Lean           & Lean is a modern, functional programming language that is designed for formalizing mathematics proof and performing auto verification. It allows users to write proofs as code, and its auto-verification ensures current written proofs are correct by verifying each logical step. Lean4 is the latest version of Lean \\

        Mathlib \& Mathlib4     & Mathlib is a large open-source library built on Lean. It contains comprehensive collection of theorems, definitions and proofs across most major areas of mathematics, including algebra, calculus and topology. Mathlib is originally build on Lean3~\cite{de2015lean}, and Mathlib4 is the Lean4 version of Mathlib. \\

        MiniF2F                 & It is a benchmark designed for evaluating automated theorem proving systems in formal mathematics. It contains a collection of mathematical problems sourced from various high-school to university-level mathematical problems. It was proposed in \citet{zheng2021minif2f}. The NL statement and proof of MiniF2F is provided by \citet{jiang2022draft}. The MiniF2F-Valid is provided in \citet{yang2024leandojo} and we provide the MiniF2F-Test. \\

        ByT5-Tacgen             & This is a tree-search Lean model provided by \citet{yang2024leandojo}, which is used to retrieve the most relevant tactic based on the current Lean status and unfinished goals. It is based on fine-tuning By-T5 using Mahtlib data \\
        \bottomrule
    \end{tabular}
    \caption{Term list}
    \label{tab:termList}
\end{table*}

\section{License}

Our dataset will be distributed under the CC BY 2.0 license, code will be distributed under the MIT license. The dataset extracted from LeanDojo~\cite{yang2024leandojo} is under CC BY 2.0. The original Mathlib4 and Lean is under Apache 2.0 license.

\end{document}